\documentstyle[aps]{revtex}
\begin{document}
\author{Kaige Wang$^{1,2}$ and Shiyao Zhu$^2$}
\address{1. CCAST (World Laboratory), P. O. Box 8730, Beijing 100080, \\
and Department of Physics, Beijing Normal University, Beijing 100875, China%
\thanks{%
Mailing address}\\
2. Department of Physics, Hong Kong Baptist University, Hong Kong, China}
\title{Two Photon Anti-Coalescence Interference: the Signature of Two Photon
Entanglement }
\maketitle

\begin{abstract}
We study a general theory on the interference of two-photon wavepacket in a
beam splitter. We find that the symmetry of two-photon spectrum plays an
important role in the manners of interference. We distinguish the
coalescence and anti-coalescence interferences, and prove that the
anti-coalescence interference is the signature of photon entanglement.

PACS number(s): 42.50.Dv, 03.65.Bz, 42.50.Ar
\end{abstract}

The two-photon interference is one of the interesting topics in quantum
optics. One of the reasons is that it concerns the nature of how two photons
interfered. Another reason could be its connection with optical quantum
information. For example, can we know photon entanglement in two-photon
interference? The first experimental observations of two-photon interference
in a beam splitter (BS) were reported in 1980s.\cite{mandel}\cite{shih1} The
key of experiment is to prepare two separated and degenerate photons with a
same frequency and polarization. This can be done by the spontaneously
parametric down-conversion (SPDC) of type I in a crystal, in which a pair of
photons, signal and idle, are produced. In the degenerate case, two photons
are incident into two ports of a 50/50 BS, no coincidence count is found at
the output ports. This effect is called the photon coalescence interference
(CI). The early theoretical explanation was based on indistinguishability of
two monochromatic single photons and it requires that two photons must meet
in BS. In the late experiments, it showed that the CI also occurs even if
two photons do not meet.\cite{chiao}\cite{shih5} In the single photon
regime, one might conclude a superluminal propagation. A succeed theoretical
explanation is to use two-photon entanglement with the help of conceptual
Feynman diagrams in which the pair of photons interfered have to be seen as
a whole, the two-photon.\cite{shih5} Just recently, Santori et al \cite{san}
has demonstrated in their experiment, that two independent single-photon
pulses emitted by a semiconductor quantum dot show a coalescence
interference in a BS. The theoretical description in this case is not
related to photon entanglement.\cite{by} A natural question could be does
the two-photon interference behave both the two-photon and the two photons
ways?

In this paper, we contribute an uniform theoretical description for
two-photon interference in BS. The theory is valid for any form of
two-photon wavepacket, entangled or un-entangled, and the coincidence
probability (CP)\ can be readily evaluated. We find that the symmetry of
two-photon spectrum plays a key role in the manners of interference. We
distinguish the coalescence and anti-coalescence interference (ACI), and
find out the necessary and sufficient condition for the perfect CI and ACI.
We prove that the photon entanglement is irrelevant to CI and necessary for
ACI. In the case of the perfect ACI, two photon state is transparent passing
the BS.

We consider a two-photon wavepacket in two orthogonal polarization modes,
designated by $H$ and $V$. The general form is described as

\begin{mathletters}
\label{1}
\begin{eqnarray}
|\Phi _{two}\rangle &=&|\Phi _{HH}\rangle +|\Phi _{VV}\rangle +|\Phi
_{HV}\rangle +|\Phi _{VH}\rangle ,  \label{1a} \\
|\Phi _{pp}\rangle &=&\sum_{\omega _1,\omega _2}C_{pp}(\omega _1,\omega
_2)a_{1p}^{\dagger }(\omega _1)a_{2p}^{\dagger }(\omega _2)|0\rangle ,\qquad
p=H,V  \label{1b} \\
|\Phi _{HV}\rangle &=&\sum_{\omega _1,\omega _2}C_{HV}(\omega _1,\omega
_2)a_{1H}^{\dagger }(\omega _1)a_{2V}^{\dagger }(\omega _2)|0\rangle ,\qquad
H\longleftrightarrow V  \label{1c}
\end{eqnarray}
in which two photons travel in two given directions, designated by
subscripts 1 and 2. In the case of only one polarization available, one has $%
|\Phi _{two}\rangle =|\Phi _{pp}\rangle $.

A beam splitter (BS) performs a linear transform for two input beams. For a
lossless BS, the general transformation between the input and output field
operators obeys\cite{cam} 
\end{mathletters}
\begin{equation}
\left( 
\begin{array}{c}
b_1 \\ 
b_2
\end{array}
\right) =S(\theta ,\phi _\tau ,\phi _\rho )\left( 
\begin{array}{c}
a_1 \\ 
a_2
\end{array}
\right) ,\qquad S(\theta ,\phi _\tau ,\phi _\rho )=\left( 
\begin{array}{cc}
e^{i\phi _\tau }\cos \theta & e^{i\phi _\rho }\sin \theta \\ 
-e^{-i\phi _\rho }\sin \theta & e^{-i\phi _\tau }\cos \theta
\end{array}
\right) ,  \label{5}
\end{equation}
where $a_i$ and $b_i$ are the field annihilation operators for the input and
output ports, respectively. The subscript $i$ ($i=1,2$) symbolizes the port
in a same propagation direction. $\theta $ characterizes both the reflection
and the transmission rates, for instance, $\theta =\pi /4$ for a 50/50 BS. $%
\phi _\tau $ and $\phi _\rho $ are two phases allowed in the unitary
transformation (\ref{5}). The corresponding transform of wavevectors in the
S-picture is written as $|\Psi \rangle _{out}=U|\Psi \rangle _{in}$ where $U$
has been given in Ref. \cite{cam}. In this paper, we use, alternatively, a
simpler method to evaluate the output wavevector. It is a duplicate of the
similar method in a dynamic quantum system: when the evolution of operators
has been known in the H-picture, one may obtain the state evolution in the
S-picture without solving the Sch\"{o}dinger equation.\cite{kaige} If the
initial state is known as $|\Psi \rangle _{in}=f(a_1,a_1^{\dagger
},a_2,a_2^{\dagger })|0\rangle $, the output state is obtained as 
\begin{equation}
|\Psi \rangle _{out}=U|\Phi \rangle _{in}=Uf(a_1,a_1^{\dagger
},a_2,a_2^{\dagger })U^{-1}U|0\rangle =f(Ua_1U^{-1},Ua_1^{\dagger
}U^{-1},Ua_2U^{-1},Ua_2^{\dagger }U^{-1})|0\rangle =f(\overline{b}_1,%
\overline{b}_1^{\dagger },\overline{b}_2,\overline{b}_2^{\dagger })|0\rangle
,  \label{6}
\end{equation}
where $\overline{b}_i\equiv Ua_iU^{-1}$ and $\overline{b}_i^{\dagger }\equiv
Ua_i^{\dagger }U^{-1}(i=1,2)$. Because $b_i=U^{-1}a_iU$ is known due to Eq. (%
\ref{5}), one may obtain its inverse transform. For a 50/50 BS, one obtains 
\begin{eqnarray}
\overline{b}_1^{\dagger }(\omega ) &=&[e^{i\phi _\tau }a_1^{\dagger }(\omega
)-e^{-i\phi _\rho }a_2^{\dagger }(\omega )]/\sqrt{2},  \label{8} \\
\overline{b}_2^{\dagger }(\omega ) &=&[e^{i\phi _\rho }a_1^{\dagger }(\omega
)+e^{-i\phi _\tau }a_2^{\dagger }(\omega )]/\sqrt{2}.  \nonumber
\end{eqnarray}
Equation (\ref{8}) is valid for a given polarization.

According to Eq. (\ref{6}), one obtains the output state for the input state
(\ref{1}) by replacing $a_j^{\dagger }$ with $\overline{b}_j^{\dagger }$.
Using Eq. (\ref{8}), for $|\Phi _{pp}\rangle $ one has

\begin{eqnarray}
|\Psi _{pp}\rangle _{out} &=&U|\Phi _{pp}\rangle =(1/2)\sum_{\omega
_1,\omega _2}C_{pp}(\omega _1,\omega _2)\{[a_{1p}^{\dagger }(\omega
_1)a_{1p}^{\dagger }(\omega _2)e^{i\phi }-  \label{11} \\
&&a_{2p}^{\dagger }(\omega _1)a_{2p}^{\dagger }(\omega _2)e^{-i\phi
}]+[a_{1p}^{\dagger }(\omega _1)a_{2p}^{\dagger }(\omega _2)-a_{2p}^{\dagger
}(\omega _1)a_{1p}^{\dagger }(\omega _2)]\}|0\rangle ,  \nonumber
\end{eqnarray}
where $\phi =\phi _\tau +\phi _\rho $. In the summation, when the frequency
variables, $\omega _1$ and $\omega _2,$ are taken in the whole frequency
space, ($\omega _1,\omega _2$) and ($\omega _2,\omega _1$) contribute to two
indistinguishable states which should be added together. For doing it, we
may take $\sum_{\omega _1,\omega _2}=\sum_{\omega _1<\omega _2}+\sum_{\omega
_1=\omega _2}+\sum_{\omega _1>\omega _2}$, and then exchange the variables $%
\omega _1$ and $\omega _2$ in the last summation. In result, Eq. (\ref{11})
is written as 
\begin{eqnarray}
|\Psi _{pp}\rangle _{out} &=&(1/2)\left( \sum_{\omega _1<\omega
_2}\{[C_{pp}(\omega _1,\omega _2)+C_{pp}(\omega _2,\omega
_1)][a_{1p}^{\dagger }(\omega _1)a_{1p}^{\dagger }(\omega _2)e^{i\phi
}-a_{2p}^{\dagger }(\omega _1)a_{2p}^{\dagger }(\omega _2)e^{-i\phi }]\right.
\label{12} \\
&&+[C_{pp}(\omega _1,\omega _2)-C_{pp}(\omega _2,\omega _1)][a_{1p}^{\dagger
}(\omega _1)a_{2p}^{\dagger }(\omega _2)-a_{1p}^{\dagger }(\omega
_2)a_{2p}^{\dagger }(\omega _1)]\}  \nonumber \\
&&\left. +\sum_\omega C_{pp}(\omega ,\omega )[(a_{1p}^{\dagger }(\omega
))^2e^{i\phi }-(a_{2p}^{\dagger }(\omega ))^2e^{-i\phi }]\right) |0\rangle
.\qquad p=H,V  \nonumber
\end{eqnarray}
In the above equation, the first and last terms describe the states in which
two photons exit from a same port, and the second term, two photons exit
from different ports, causing a ''click-click'' in the coincidence
measurement. Because the output states corresponding to $|\Phi _{HV}\rangle $
and $|\Phi _{VH}\rangle $ exist indistinguishable states and should be
considered in a coherent superposition, one has

\begin{eqnarray}
|\Psi _{HV+VH}\rangle _{out} &=&U(|\Phi _{HV}\rangle +|\Phi _{VH}\rangle )
\label{13} \\
&=&(1/2)\sum_{\omega _1,\omega _2}\{[C_{HV}(\omega _1,\omega
_2)+C_{VH}(\omega _2,\omega _1)][a_{1H}^{\dagger }(\omega _1)a_{1V}^{\dagger
}(\omega _2)e^{i\phi }-a_{2H}^{\dagger }(\omega _1)a_{2V}^{\dagger }(\omega
_2)e^{-i\phi }]  \nonumber \\
&&+[C_{HV}(\omega _1,\omega _2)-C_{VH}(\omega _2,\omega _1)][a_{1H}^{\dagger
}(\omega _1)a_{2V}^{\dagger }(\omega _2)-a_{2H}^{\dagger }(\omega
_1)a_{1V}^{\dagger }(\omega _2)]\}|0\rangle .\qquad H\longleftrightarrow V 
\nonumber
\end{eqnarray}
Again, the first term and the second term describe the states in which two
photons exit from a same port and different ports, respectively. Note that
there is no interference among the states $|\Phi _{HH}\rangle ,|\Phi
_{VV}\rangle $ and $(|\Phi _{HV}\rangle +|\Phi _{VH}\rangle )$, since the
two-photon polarization configurations are distinguishable.

Equations (\ref{12}) and (\ref{13}) have implied that the symmetry of the
spectra plays important role to the manners of interference. The symmetric
spectrum reads as

\begin{mathletters}
\label{14s}
\begin{eqnarray}
C_{pp}(\omega _1,\omega _2) &\equiv &C_{pp}(\omega _2,\omega _1),\qquad
p=H,V,  \label{14as} \\
C_{HV}(\omega _1,\omega _2) &\equiv &C_{VH}(\omega _2,\omega _1).
\label{14bs}
\end{eqnarray}
In this case, the second terms in Eqs. (\ref{12}) and (\ref{13}) vanish. It
causes so called perfectly ''coalescence interference'' (CI), that is, two
photons coalesce together and cause a null coincidence probability. This
effect has been paid more attention in the literatures.\cite{mandel}\cite
{shih1} On the contrary, the anti-symmetric spectrum satisfies condition 
\end{mathletters}
\begin{mathletters}
\label{14}
\begin{eqnarray}
C_{pp}(\omega _1,\omega _2) &\equiv &-C_{pp}(\omega _2,\omega _1),\qquad
p=H,V,  \label{14a} \\
C_{HV}(\omega _1,\omega _2) &\equiv &-C_{VH}(\omega _2,\omega _1),
\label{14b}
\end{eqnarray}
under which two output photons never go together and the CP is unit. We call
it the perfect anti-coalescence interference (ACI). Precisely, Eqs. (\ref{12}%
) and (\ref{13}) are respectively reduced to 
\end{mathletters}
\begin{eqnarray}
|\Psi _{pp}\rangle _{out} &=&\sum_{\omega _1<\omega _2}C_{pp}(\omega
_1,\omega _2)[a_{1p}^{\dagger }(\omega _1)a_{2p}^{\dagger }(\omega
_2)-a_{1p}^{\dagger }(\omega _2)a_{2p}^{\dagger }(\omega _1)]|0\rangle
\label{15} \\
&=&\sum_{\omega _1,\omega _2}C_{pp}(\omega _1,\omega _2)a_{1p}^{\dagger
}(\omega _1)a_{2p}^{\dagger }(\omega _2)|0\rangle =|\Phi _{pp}\rangle , 
\nonumber
\end{eqnarray}
and 
\begin{eqnarray}
|\Psi _{HV}\rangle _{out} &=&\sum_{\omega _1,\omega _2}C_{HV}(\omega
_1,\omega _2)[a_{1H}^{\dagger }(\omega _1)a_{2V}^{\dagger }(\omega
_2)-a_{2H}^{\dagger }(\omega _1)a_{1V}^{\dagger }(\omega _2)]|0\rangle
\label{16} \\
&=&\sum_{\omega _1,\omega _2}[C_{HV}(\omega _1,\omega _2)a_{1H}^{\dagger
}(\omega _1)a_{2V}^{\dagger }(\omega _2)+C_{VH}(\omega _2,\omega
_1)a_{2H}^{\dagger }(\omega _1)a_{1V}^{\dagger }(\omega _2)]|0\rangle =|\Phi
_{HV}\rangle .  \nonumber
\end{eqnarray}
That is, the output state are identical to the input. Physically, this
two-photon wavepacket is perfectly transparent passing BS. We would like to
emphasize that conditions (\ref{14a}) and (\ref{14b}) are sufficient and
necessary for Eqs. (\ref{15}) and (\ref{16}), respectively.

In general, we may measure the coincidence probability (CP) to evaluate the
CI\ or ACI effects. According to Eqs. (\ref{12}) and (\ref{13}), the CP is
obtained as 
\begin{equation}
P_c=\frac 12\left\{ 1-\frac 12\int_{-\infty }^\infty d\omega _1\int_{-\infty
}^\infty d\omega _2[2C_{HV}(\omega _1,\omega _2)C_{VH}^{*}(\omega _2,\omega
_1)+\sum_{p=H,V}C_{pp}(\omega _1,\omega _2)C_{pp}^{*}(\omega _2,\omega _1)+%
\text{c.c.}]\right\} ,  \label{17}
\end{equation}
where the three integral terms which may increase or decrease the
coincidence indicate the two-photon interference. When all the interference
terms vanish, there is no two-photon interference and we have $P_c=1/2.$ $%
P_c<1/2$ and $P_c>1/2$ stand for CI and ACI, respectively, When the
anti-symmetric condition (\ref{14}) is satisfied, $P_c=1$, which witnesses
the two-photon transparent state.

The CI can occur for both the entangled and the un-entangled two photon
wavepackets, whereas ACI never occurs for the un-entangled two photon state.
Assume two independent single-photon wavepackets being in two orthogonal
polarizations, each of them is written as $|\Phi _{one}\rangle
_j=\sum_\omega [C_{jH}(\omega )a_{jH}^{\dagger }(\omega )+C_{jV}(\omega
)a_{jV}^{\dagger }(\omega )]|0\rangle ,$ ($j=1,2$). The coincidence
probability is 
\begin{equation}
P_c=\frac 12\left[ 1-\left| \int_{-\infty }^\infty d\omega [C_{1H}(\omega
)C_{2H}^{*}(\omega )+C_{1V}(\omega )C_{2V}^{*}(\omega )]\right| ^2\right]
\leq \frac 12.  \label{35}
\end{equation}
Therefore, ACI effect never occurs for un-entangled two photon wavepacket.

In the following examples, we show how to demonstrate two photon
entanglement in experimental observations. First, we consider the case of
one polarization. A two-photon wavepacket is assumed in a symmetric form 
\begin{equation}
Q(\omega _1,\omega _2)=g(\omega _1+\omega _2-2\Omega )f(\omega _1-\Omega
)f(\omega _2-\Omega ),  \label{20}
\end{equation}
where $\Omega $ is the central frequency for both beams, and $f$ describes
the spectral profile for each single-photon. In the SPDC of type I, a pair
of converted photons can be in this form\cite{shih0} However, Eq. (\ref{20})
can describe two independent single-photon wavepackets either, provided $%
g(x)=1$. Because of the symmetry of the two-photon spectrum $Q(\omega
_1,\omega _2)$, the perfect CI must occur. In the experiment, beam $j$
travels a path $z_j$, and enters the input port of a BS. So that the two
photon spectrum at the input ports is $C(\omega _1,\omega _2)=Q(\omega
_1,\omega _2)\exp [i(\omega _1z_1+\omega _2z_2)/c]$. Assume the
single-photon spectrum is a Gaussian $f(x)=\exp [-x^2/(2\sigma ^2)]$ with a
bandwidth $\sigma $, we may obtain the coincidence probability $%
P_c=(1/2)\{1-\exp [-(\sigma \Delta z/c)^2/2]\}$where $\Delta z=z_2-z_1$. It
shows a famous interference dip.\cite{mandel}\cite{shih1}\cite{san} This
result is independent of the form of function $g(x)$, whether in
entanglement or not. Therefore, our theory presents an uniform explanation
to the ''dip''.

Nevertheless, the dip can not tell us the entanglement between two photons.
In order to show the entanglement, one may introduce an additional phase in
two-photon spectrum for changing its symmetry. The method was reported in
Refs. \cite{chiao}\cite{shih5}, in which the authors wanted to show that two
photon interference occurs even if they ''do not meet''. Let beam 1 splits
two parts, one travels a short path $L_s$ and the other, a long path $L_l$.
Then these two sub-beams incorporate a beam again which interferes with beam
2 traveling a path $z_2$. The new two-photon spectrum at the input ports of
BS is obtained as 
\begin{equation}
C(\omega _1,\omega _2)=Q(\omega _1,\omega _2)[e^{i\omega _1L_l/c}+e^{i\omega
_1L_s/c}]e^{i\omega _2z_2}=2Q(\omega _1,\omega _2)e^{i(\omega _1z_1+\omega
_2z_2)/c}\cos (\omega _1\Delta L/c),  \label{23}
\end{equation}
where $z_1=(L_l+L_s)/2$ and $\Delta L=(L_l-L_s)/2$. Substituting the
symmetric spectrum (\ref{20}) into Eq. (\ref{23}), one obtains 
\begin{equation}
C(\nu _1,\nu _2)=2e^{i\Omega (z_1+z_2)/c}g(\nu _1+\nu _2)f(\nu _1)f(\nu
_2)e^{i(\nu _1z_1+\nu _2z_2)/c}\cos (\nu _1\Delta L/c+\alpha ),  \label{24}
\end{equation}
where $\nu _i=\omega _i-\Omega ,$ $(i=1,2)$ and $\alpha \equiv \Omega \Delta
L/c$. In the case $g(x)\rightarrow \delta (x),$ which denotes a perfect
phase matching in SPDC, at the balanced position $z_1=z_2$, the above
spectrum is symmetric (anti-symmetric) when $\alpha =n\pi $ ($\alpha
=[n+(1/2)]\pi $). Let $g(x)=\exp [-x^2/(2\sigma _p^2)]$, we can calculate
exactly the CP 
\begin{equation}
P_c=\frac 12\{1-\frac 1{2B}[\cos (2\alpha )e^{-\frac 12[\frac{\beta ^2}{%
2+\beta ^2}\Delta L^2+\Delta z^2](\frac \sigma c)^2}+\frac 12e^{-\frac 12%
(\Delta L+\Delta z)^2(\frac \sigma c)^2}+\frac 12e^{-\frac 12(\Delta
L-\Delta z)^2(\frac \sigma c)^2}]\},  \label{18}
\end{equation}
where $B=\frac 12[1+\cos (2\alpha )\exp [-\frac{1+\beta ^2}{2+\beta ^2}%
\Delta L^2(\frac \sigma c)^2]$ and $\beta \equiv \sigma _p/\sigma $. Figure
1 shows the CP versus $\Delta z(\sigma /c)$ for two-photon state (\ref{24}).
In Fig. 1a, when $\alpha =$ $n\pi $ and $\alpha =[n+(1/2)]\pi ,$ the CP
shows the CI (dashed lines) and ACI (solid lines) effects, respectively. For 
$\beta =0$, $g(x)\rightarrow \delta (x)$, both CI\ and ACI are perfect.
However, for $\beta =1$ $\rightarrow \infty $, all the CP curves are almost
superposed by showing no interference at the balanced position. For two
independent single-photon wavepackets ($\beta =\infty $ ), when $\Delta
L(\sigma /c)$ is smaller, the CI may appear at the balanced position but the
ACI never occurs, as shown in Fig. 1b.

In the next example, we consider two photons being in two orthogonal
polarizations. Assume a single-photon state produced by a single-photon
source $|\Phi _{one}\rangle =\sum_\omega f(\omega -\Omega )[a_H^{\dagger
}(\omega )+a_V^{\dagger }(\omega )]|0\rangle $, one sets up an experiment
for two independent single-photon wavepackets interfered in a BS. Similarly,
in the experiment, beam $j$ travels a path $z_j$, before entering BS. In
order to show a different manner of interference, one puts a wave-plate in
path 1 which introduces an additional relative phase $\alpha $ to a
particular polarization.\cite{zei} The two-photon state for these two
independent wavepackets is written as 
\begin{equation}
\sum_{\omega _1,\omega _2}f(\omega _1-\Omega )f(\omega _2-\Omega
)e^{i(\omega _1z_1+\omega _2z_2)/c}[a_{1H}^{\dagger }(\omega _1)+e^{i\alpha
}a_{1V}^{\dagger }(\omega _1)][a_{2H}^{\dagger }(\omega _2)+a_{2V}^{\dagger
}(\omega _2)]|0\rangle .  \label{19}
\end{equation}
If $f(x)=\exp [-x^2/(2\sigma ^2)]$, the CP can be calculated as 
\begin{equation}
P_c=(1/2)[1-(1/2)(1+\cos \alpha )e^{-\frac 12(\Delta z\cdot \sigma /c)^2}].
\label{21}
\end{equation}
On the other hand, for SPDC of type II, the two photon state with
polarization entanglement can be generated in the overlap between the o-ray
cone and the e-ray cone.\cite{zei} The two-photon wavepacket with the
polarization entanglement can be described as 
\begin{equation}
\sum_{\omega _1,\omega _2}Q(\omega _1,\omega _2)e^{i(\omega _1z_1+\omega
_2z_2)/c}[a_{1H}^{\dagger }(\omega _1)a_{2V}^{\dagger }(\omega
_2)+e^{i\alpha }a_{1V}^{\dagger }(\omega _1)a_{2H}^{\dagger }(\omega
_2)]|0\rangle ,  \label{22}
\end{equation}
where $Q(\omega _1,\omega _2)$ has been defined in Eq. (\ref{20}). The phase 
$\alpha $ depends on the crystal birefringence, but, as mentioned above, it
can be set as desired.\cite{zei} For $\alpha =0$ and $\alpha =\pi $, at the
balanced case $z_1=z_2,$ Eq. (\ref{22}) satisfies the symmetric and
anti-symmetric conditions, (\ref{14bs}) and (\ref{14b}), respectively. Let $%
f(x)$ is a Gaussian, the coincidence probability is obtained as 
\begin{equation}
P_c=(1/2)(1-\cos \alpha \cdot e^{-\frac 12\sigma ^2\Delta z^2/c^2}).
\label{29}
\end{equation}
Not that the result is independent of the function form $g(x)$, so long as
it is symmetric to $\omega _1$ and $\omega _2$. Figures 2a and 2b show the
CP versus the normalized path difference for un-entangled and entangled
wavepackets, (\ref{19}) and (\ref{22}), respectively. Obviously, the
interference pattern depends on the phase $\alpha $. For $\alpha =0,$ the
both cases show the perfect CI effect. On the contrary, for $\alpha =\pi ,$
the entangled two-photon wavepacket shows the perfect ACI, whereas the two
independent single-photon wavepackets show ''no interference''. Precisely,
in the late case, the CP introduced by three interference terms in Eq. (\ref
{17}) is cancelled exactly. However, for $\alpha =\pi /2,$ there is no
interference for the entangled wavepacket because of the out of phase of the
probability amplitudes.

In conclusion, we have shown that, in the quantum state language, two-photon
interference originates from indistinguishability of two-photon states,
instead of two photons. This interference mechanism is true for both
entangled and un-entangled two-photon wavepackets. The photon entanglement
may affect the manners of interference. We propose that ACI effect is the
signature of two-photon entanglement, which might be a simpler way to
witness the entanglement, besides the violation of Bell's inequality.

The authors thank G. X. Li for stimulating discussions. This research was
supported by the National Program of Fundamental Research No. 2001CB309310
and the National Natural Science Foundation of China, Project Nos. 10074008
and 60278021, and also supported by FRG from HKBU and RGC from HK Government.

\bigskip\ captions of figures:

Fig.1 In the case of one polarization mode, coincidence probability versus
the normalized path difference $\Delta z(\sigma /c)$. (a) for $\Delta
L(\sigma /c)=5$, $\alpha =n\pi $ (dashed) and $(n+1/2)\pi $ (solid), curves
1,2,3,4 and 5 are for $\beta =0,0.2,0.5,1$ and $\infty $, respectively; (b)
for $\beta =\infty $ and $\Delta L(\sigma /c)=1$, dashed, dotted and solid
curves are for $\alpha =n\pi $, $(n+1/4)\pi $ and $(n+1/2)\pi $,
respectively.

Fig. 2 In the case of two orthogonal polarization modes, coincidence
probability versus the normalized path difference $\Delta z(\sigma /c)$ for $%
\alpha =0,\pi /2,$ and $\pi $; (a) two independent single-photon
wavepackets, and (b) two photon wavepacket with polarization entanglement.

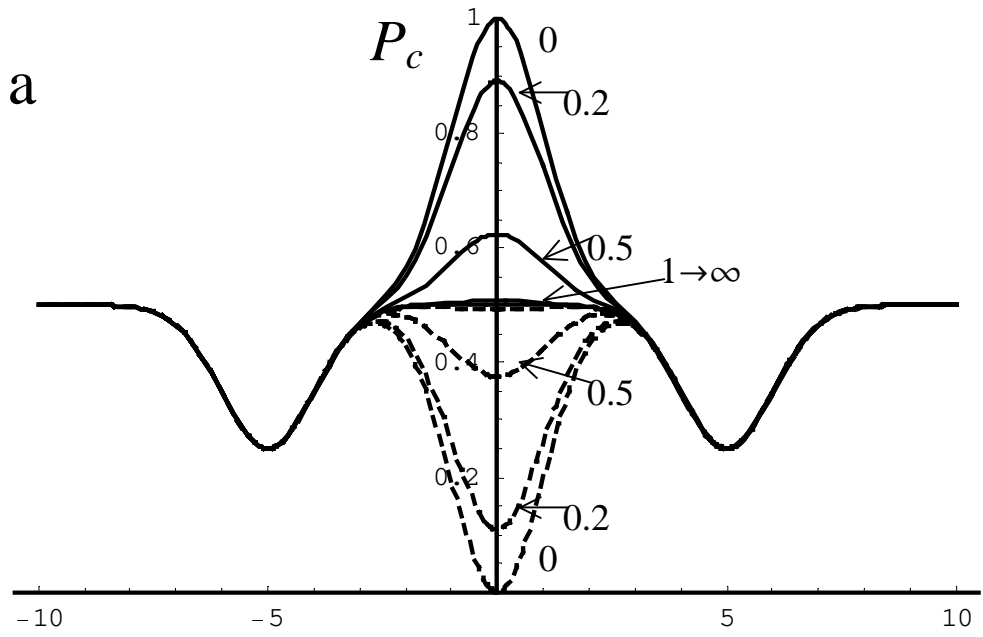
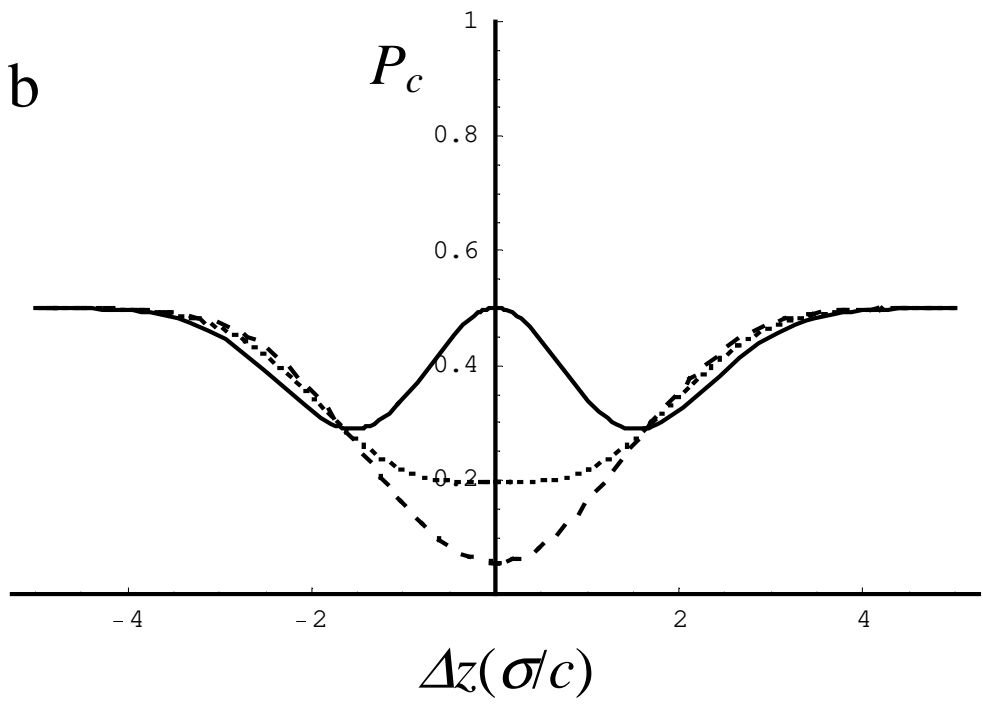
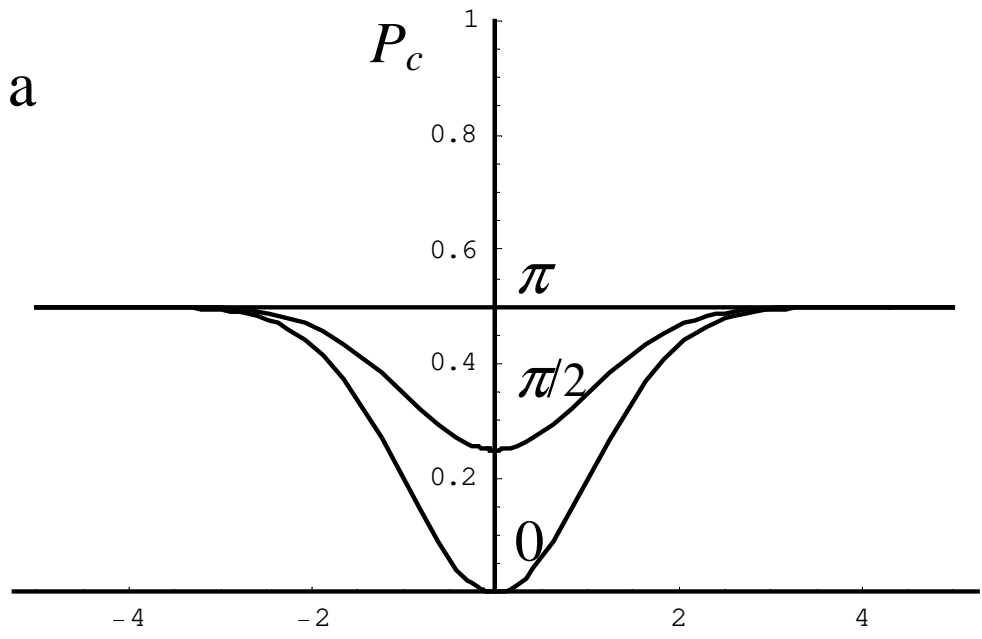
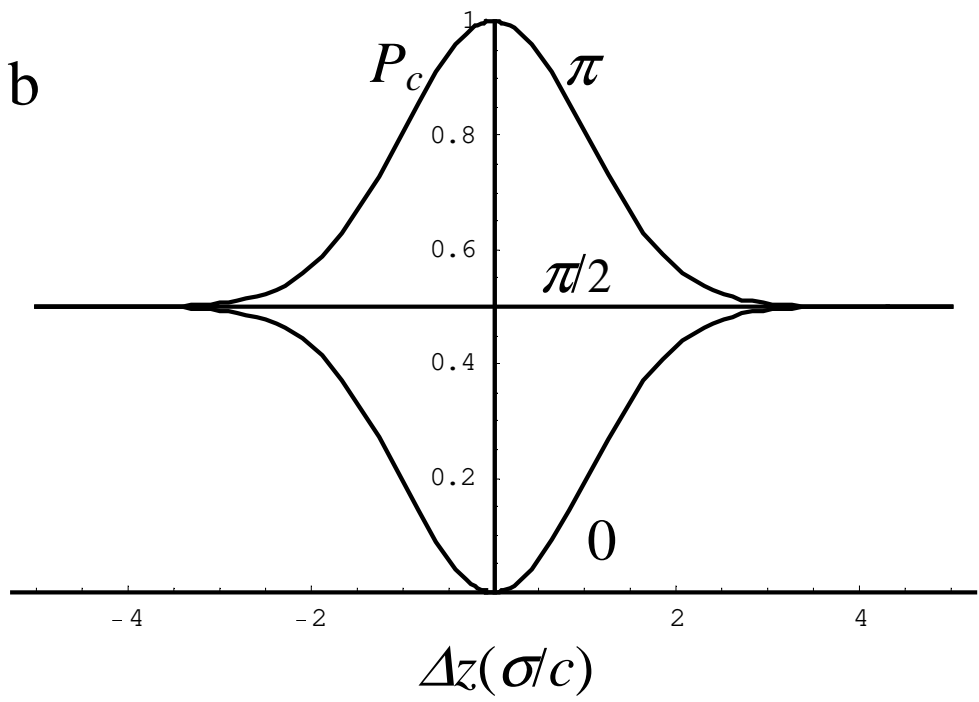
\end{document}